\documentstyle[11pt,newpasp,twoside,psfig]{article}
\begin{document}

\title{The production mechanism of radio jets in AGN and quasar grand 
unification}
\author{Mark Lacy}
\affil{SIRTF Science Center, California Institute of Technology, 1200 E.\ California Boulevard, Pasadena, CA 91125}

\setcounter{page}{1}
\index{Lacy, M.}

\begin{abstract}
Recent advances in estimating black hole masses for AGN show that 
radio luminosity is dependent on black hole mass and accretion rate. 
In this paper we outline a possible scheme for unifying radio-quiet and 
radio-loud AGN. We take the ``optimistic'' view that the mass and spin of 
the central black hole, the accretion rate onto it,
plus orientation and a weak environmental dependence, fully determine the 
observed properties of AGN.
\end{abstract}

\paragraph{The production of powerful radio jets}

There is good observational evidence that radio jet power is closely linked
to both the mass of the black hole and its accretion rate. Links between 
the mass of the black hole and its radio luminosity have been found in both
radio-weak elliptical galaxies (Franceschini, Vercellone \& Fabian 1998) 
and quasars (Laor 2000; Lacy et al.\ 2001; Boroson 2002).
Essentially all steep-spectrum radio-selected quasars and BLLac 
objects have black holes more massive than $\sim 10^8M_{\odot}$ (e.g.\ 
McLure \& Jarvis 2002; Barth, Ho \& Sargent 2002; O'Dowd, Urry \& Scarpa 
2002), and all powerful radio galaxies exist in massive hosts, almost 
invariably giant ellipticals. Recent claims that a significant fraction of 
flat-spectrum radio-luminous quasars have black holes $\sim 10^{6-7} 
M_{\odot}$ (Oshlack, Webster \& Whiting 2002), have been questioned by 
Jarvis \& McLure (2002) on the basis of orientation biases.

Radio luminosity is roughly linearly
correlated with accretion rate for quasars (e.g.\ Lacy et al.\ 2001). 
At low accretion rates (in the ADAF regime), however, black holes seem to be 
able to deliver more radio jet power than expected if the linear correlation
continues to low accretion rates, i.e.\ low 
luminosity AGN tend to be radio louder than most quasars (Ho 2002).

Most models predict than black hole 
spin should be important for determining radio jet luminosity.
Indeed, the residual scatter of radio luminosity about the best-fit 
combination of black hole mass and accretion rate (a radio ``fundamental 
plane'') is 1-2 orders of magnitude, and 
probably not just due to measurement error, suggesting that spin, and/or
some other parameter(s) need to be taken into account. Recent 
results from a study of the Fe K$\alpha$ line in MCG-6-30-15 
indicates that the black hole
in this radio-quiet Seyfert galaxy may be spinning rapidly, however 
(Fabian, these proceedings).

\paragraph{Broad absorption line quasars}

The fraction of highly radio-luminous BALQs is small compared to that in 
samples of radio-quiet quasars. If broad absorption lines are only seen in 
objects accreting close to the Eddington rate (Boroson 2002), then this 
might be explained as a selection effect. Most radio-loud quasars have high 
black hole masses, thus a radio-loud quasar accreting at near the Eddington 
rate would be highly luminous, and hence very rare.

\bigskip

\begin{figure}[t]
\centerline{
\plotfiddle{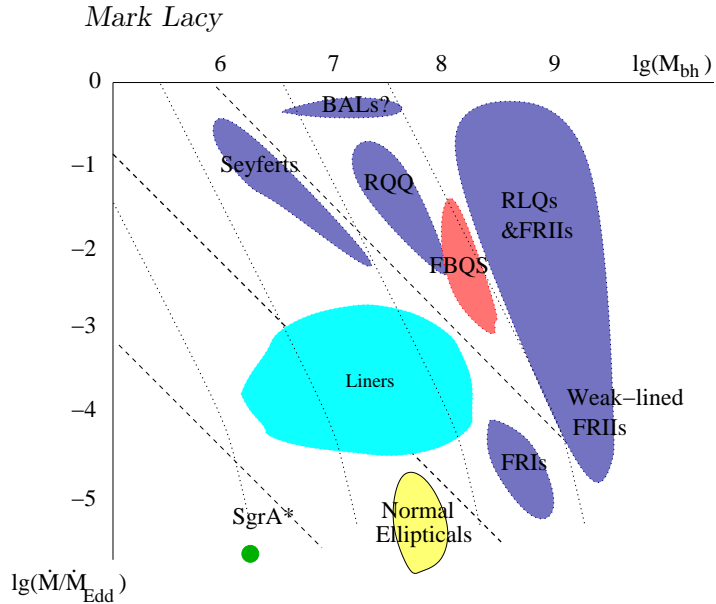}{200pt}{0}{50}{50}{-330pt}{0pt}
}
\caption{A tentative AGN grand unification scheme. 
Orientation should be thought of as another, orthogonal axis in this 
plot. Lines of constant radio luminosity are shown dotted, lines of constant
accretion (optical) luminosity are shown dashed.}
\label{fig1}
\end{figure}


\acknowledgements
I thank Aaron Barth and Mike Brotherton for helpful discussions.
This work was carried out at the Jet Propulsion Laboratory, California 
Institute of Technology, under contract with NASA.

\end{document}